\documentclass[twocolumn]{aastex62}
\usepackage[version=4]{mhchem}
\usepackage{breqn}
\usepackage{amsmath}

\newcommand{\kms}{$\rm km~s^{-1}$}

\shorttitle{Orion source I ionized outflow?}
\shortauthors{Wright et al.}
\begin{document}
\title{\bf An ionized outflow in Orion-KL source I?}
%\title{A Binary Protostar in Orion ?}
%\title{Observations of Orion Source I Disk and Outflow Interface}
%\title{Orion Source I as the disk driven outflow paradigm}
%% The \author command is the same as before except it now takes an optional
%% arguement which is the 16 digit ORCID. The syntax is:
%% \author[xxxx-xxxx-xxxx-xxxx]{Author Name}
%%
%% This will hyperlink the author name to the author's ORCID page. Note that
%% during compilation, LaTeX will do some limited checking of the format of
%% the ID to make sure it is valid.
%%
%% Use \affiliation for affiliation information. The old \affil is now aliased
%% to \affiliation. AASTeX v6.2 will automatically index these in the header.
%% When a duplicate is found its index will be the same as its previous entry.
%%

%% Use \email to set provide email addresses. Each \email will appear on its
%% own line so you can put multiple email address in one \email call. A new
%% \correspondingauthor command is available in V6.2 to identify the
%% corresponding author of the manuscript. It is the author's responsibility
%% to make sure this name is also in the author list.
%%
%% While authors can be grouped inside the same \author and \affiliation
%% commands it is better to have a single author for each. This allows for
%% one to exploit all the new benefits and should make book-keeping easier.
%%
%% If done correctly the peer review system will be able to
%% automatically put the author and affiliation information from the manuscript
%% and save the corresponding author the trouble of entering it by hand.

\correspondingauthor{Melvyn Wright}
\email{melvyn@berkeley.edu}

\author{Melvyn Wright}
\affiliation{Department of Astronomy, University of California, 501 Campbell Hall, Berkeley CA 94720-3441, USA}

\author{Tomoya Hirota}
\affiliation{Mizusawa VLBI Observatory, National Astronomical Observatory of Japan, 
2-12, Hoshigaoka, Mizusawa, Oshu, Iwate 023-0861, Japan}

\author{Jan Forbrich}
\affiliation{Centre for Astrophysics Research, University of Hertfordshire, College Lane, Hatfield AL10 9AB, UK}

\author[0000-0001-6765-9609]{Richard Plambeck}
\affiliation{Radio Astronomy Lab, University of California, 501 Campbell Hall, Berkeley CA 94720-3441, USA}

\author{John Bally}
\affil{CASA, University of Colorado, 389-UCB, Boulder, CO 80309, USA}

\author[0000-0002-2542-7743]{Ciriaco Goddi}
\affil{
Universidade de São Paulo, Instituto de Astronomia, Geofísica e Ciências Atmosféricas, Departamento de Astronomia, São Paulo, SP 05508-090, Brazil}
\affil{
Dipartimento di Fisica, Università degli Studi di Cagliari, SP Monserrato-Sestu km 0.7, I-09042 Monserrato, Italy}
\affil{
INAF - Osservatorio Astronomico di Cagliari, via della Scienza 5, I-09047 Selargius (CA), Italy}
\affil{
INFN, Sezione di Cagliari, Cittadella Univ., I-09042 Monserrato (CA), Italy}

\author[0000-0001-6431-9633]{Adam Ginsburg}
\affil{Department of Astronomy, University of Florida
211 Bryant Space Science Center
P.O Box 112055, Gainesville, FL 32611-2055 USA}

\author[0000-0003-1254-4817]{Brett A. McGuire}
\affil{Department of Chemistry, Massachusetts Institute of Technology, Cambridge, MA 02139, USA}
\affil{National Radio Astronomy Observatory, Charlottesville, VA 22903, USA}

%\author{etal.}

 \begin{abstract}
 
 We present images at 6 and 14~GHz of Source I in Orion-KL.  At higher frequencies, from 43 to 340 GHz, images of this source are dominated by thermal emission from dust in a 100 AU diameter circumstellar disk, but at 6 and 14 GHz the emission is elongated along the minor axis of the disk, aligned with the SiO bipolar outflow from the central object.  Gaussian fits to the 6, 14, 43, and 99 GHz images find a component along the disk minor axis whose flux and length vary with frequency consistent with free-free emission from an ionized outflow.   The data favor a broad outflow from a disk wind, rather than a narrow ionized jet.   Source I was undetected in higher resolution 5 GHz e-MERLIN observations obtained in 2021. 
The 5-6 GHz structure of SrcI may be resolved out by the high sidelobe structure of the e-MERLIN synthesized beam, or  be time variable.

\end{abstract}


%% tied to the reference list via symbolic KEYs. The KEY corresponds to the KEY in the \bibitem in the reference list below. 

{\bf \section{Introduction} \label{sec:intro}}
 Jets and outflows are closely associated with accretion in the formation of stars.  Free–free (thermal) radio continuum emission  is associated with these jets. The emission is weak and compact, and high angular resolution at centimeter wavelengths is required map this emission
because of high opacity at millimeter and IR wavelengths, e.g. review by \citet{Anglada2018}.

 Protostellar outflows transfer momentum and kinetic energy to outer envelopes and influence future star-formation in surrounding regions. Outflows extract excess angular momenta from protostars and their surrounding disks to enable mass accretion. The mechanisms to drive outflows, such as entrainment of molecular gas by a higher-velocity jet, or an MHD-driven outflow by magneto-centrifugal disk winds are still poorly understood. 
 \citet{Commercon2022}
 studied the evolution of disks, accretion and outflow in massive stars using high-resolution numerical simulations with several different numerical codes. Outflows may be driven by both radiation pressure and MHD mechanisms. The properties of the accretion disk around massive protostars depends on the physics, including  hydrodynamics, magnetic fields, and ambipolar diffusion.
 In this paper we present new observations of the
 outflow and accretion disk in the massive young protostar, Source I (SrcI), and discuss the possible emission mechanisms which may help to constrain the theoretical models of massive star formation.
 
The Kleinmann-Low Nebula in Orion, at a distance 415 pc   \citep{Menten2007,Kim2008,Kounkel2018}, is the nearest HII region 
in which massive ($M >8$~$M_{\odot}$) stars are forming. The two most massive objects in this region, Source~I and the Becklin-Neugebauer Object (BN), appear to be recoiling from one another at 35-40~\kms\ \citep{Rodriguez2005, Gomez2008, Goddi2011}, suggesting that they were ejected from a multiple system via dynamical decay approximately 550 years ago
\citep{Rodriguez2017, Rodriguez2020, Bally2020}. 

SrcI has an estimated mass of $\sim$15~$M_{\odot}$ \citep{Ginsburg2018}, with a rotating accretion disk, a hot inner core obscured by a dusty outer envelope, and a young molecular outflow that is prominent in shock-tracing SiO in several rotational-vibration levels extending out along the minor axis of the disk.
 
The region around SrcI has been well studied since the 1980s
\citep{Hirota2014, Plambeck2016}; it is an active source associated with variable SiO and \ce{H2O} masers \citep{Reid2007, Goddi2009, Plambeck2009, Matthews2010, Goddi2011,  Niederhofer2012, Greenhill2013}.
\citet{Ginsburg2018} fitted the observed structure of the disk using 90 and 230~GHz ALMA continuum observations at 50 and 20 mas resolution. They determined that the disk has a length of $\sim$100 AU, and vertical FWHM height of $\sim$20 AU. 
The disk major axis increases with observing frequency from 43 to 340~GHz which is expected as the dust optical depth increases.
Images of the spectral index distributions from 43 to 340~GHz show an extensive region with spectral index $<$ 2 along the minor axis of the disk. Although the high opacity at 220 - 340 GHz hides the internal structure of the disk, images at 43 to 99 GHz reveal structure within the disk \citep{Wright2022}.
At lower frequencies, the spectral energy distribution suggests that free-free emission may become important.
\citet{Plambeck2016} analysed the continuum flux densities measured for SrcI from 4 to 690 GHz. Free–free emission may become important where the dust emission becomes optically thin. 

In this paper, we present 6 and 14~GHz images of emission the SrcI disk in Orion-KL with $\sim 0.3'' \times 0.2''$ resolution,
and observations with the e-MERLIN telescope at 5 GHz with a synthesized beam FWHM {120}$\times$30 mas.

We analyse the continuum images from 6 to 99~GHz to trace the structure along the minor axis of the Source I disk.
The 6 and 14~GHz emission extends out along the minor axis of the disk, and may be evidence for a dusty ionized outflow from SrcI.

The paper is organized as follows:  \S2 presents the observations and data reduction, \S3 discusses the results for the SrcI disk and outflow,
and discusses different emission mechanisms,
\S4 summarizes the conclusions.

\medskip
{\bf \section{Observations and Data Reduction}}

We used JVLA and ALMA data at 6, 14, 43, 86, 99, 223, and 340 GHz to image the continuum emission from SrcI.
Table~1 provides a summary of the observations, including project codes and synthesized beam FWHM obtained with robust= {-2} weighting of the uv-data.The \textsc{Miriad} software package \citep{Sault1995} was used to analyze the data in this paper.  

The 43~GHz and 99~GHz (ALMA Band 3; B3) observations and calibration are described in \citet{Hirota2020} and \citet{Wright2020}. The 86 GHz observations are described in \citet{Wright2022}.
The 223~GHz (ALMA Band 6; B6) and 340~GHz (ALMA Band 7; B7) observations and calibration are described in \citet{Ginsburg2018}.

\begin{figure}
\includegraphics[width=0.5\textwidth, clip, trim=5cm 4cm 2.5cm 1cm]{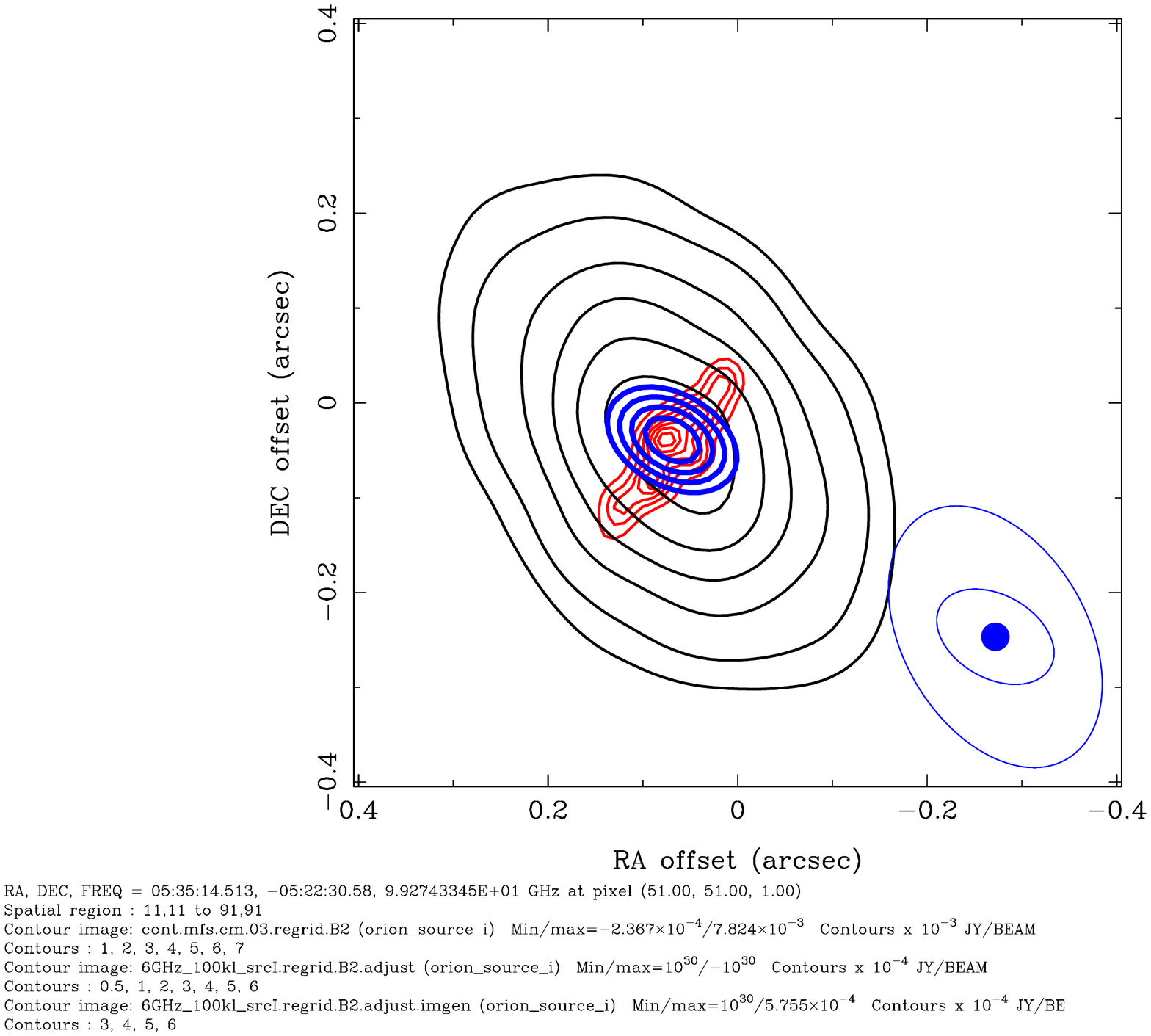}
\includegraphics[width=0.5\textwidth, clip, trim=5cm 4cm 2.5cm 1cm]{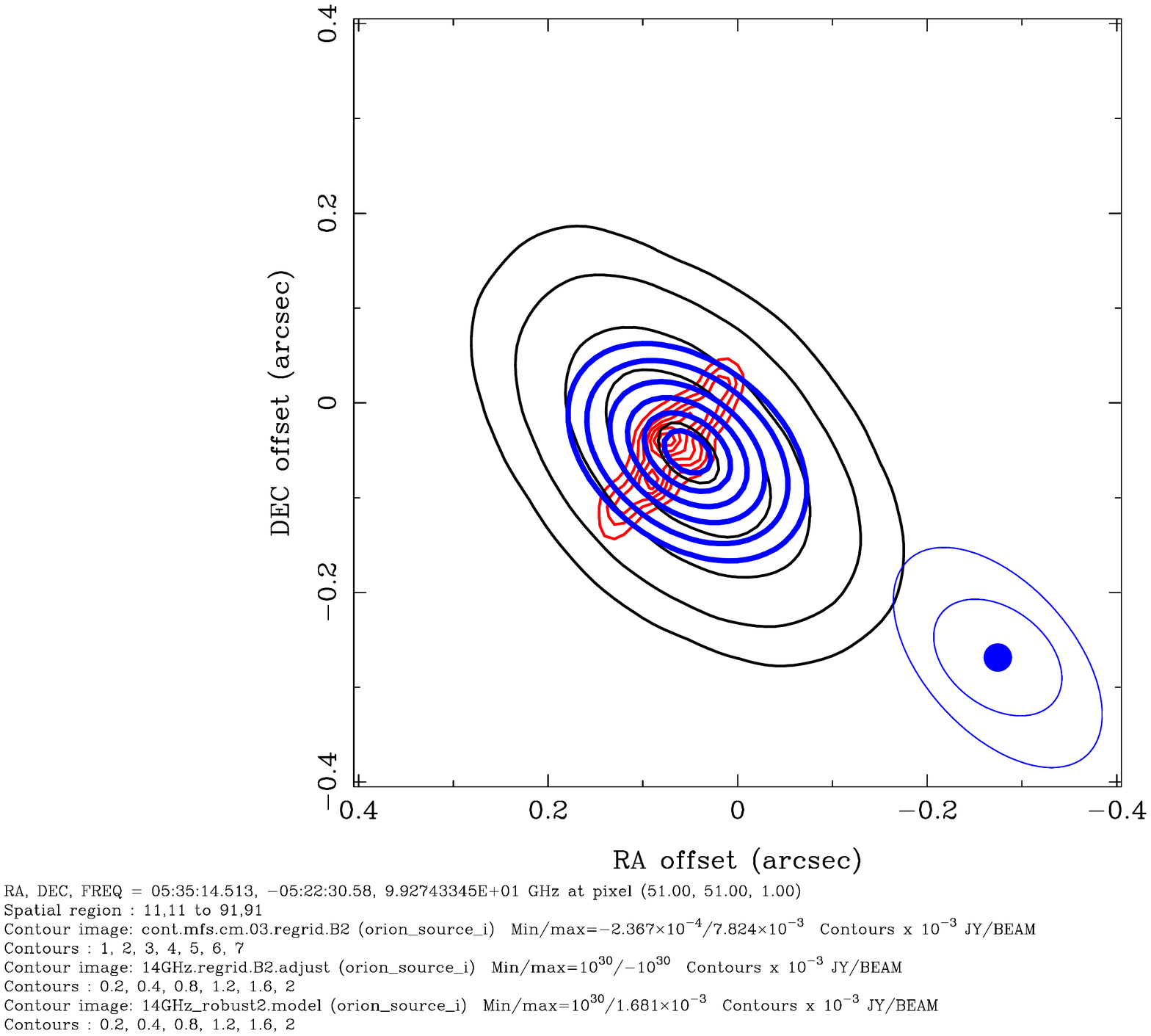}
\caption{
\textbf{(Top)} Source I continuum emission at 6~GHz (black), and Gaussian model (blue).
 Contour levels: 0.05 0.1 0.2 0.3 0.4 0.5 0.6 mJy/beam,
 superposed on the 99 GHz model showing the disk emission at 30 mas resolution (red).
 Contour levels 1 2 3 4 5 6 7 mJy/beam.
The synthesized beam and the Gaussian model at 6 GHz, and the 99 GHz model FWHM are shown as ellipses in the lower right.
\textbf{(Bottom)} Source I continuum emission at 14~GHz (black), and Gaussian model (blue).
 Contour levels:  0.2 0.4 0.8 1.2 1.6 2 mJy/beam,
 superposed on the 99 GHz model showing the disk emission at 30 mas resolution (red).
 Contour levels 1 2 3 4 5 6 7 mJy/beam.
The synthesized beam and the Gaussian model at 14 GHz, and the 99 GHz model FWHM are shown as ellipses in the lower right.
\label{fig:6GHz_image}}
\end{figure}

We present a continuum image at 6 GHz using 
the data described in \citet{Forbrich2016}.
Figure~\ref{fig:6GHz_image} shows the image of SrcI using uv-data $>100 k\lambda $ to filter out large scale structure from Orion.  
Observations at 14 GHz were obtained using the JVLA on 2022-06-04. These observations were calibrated using observatory-provided scripts. Synthesized images were made using the \textsc{casa} task tclean.
The synthesised beam FWHM was 280 $\times$ 160 mas in PA 42\degr ~using robust= {2} weighting of the uv-data. Figure~\ref{fig:6GHz_image} shows the resulting image of SrcI.

Observations with the %{\bf{\textcolor{red}
{e-MERLIN} 
array were obtained 2021-10-23, -25, -26 (epoch 2021.81) at 5 GHz
Six antennas among 7 e-MERLIN stations (except Darnhall) were used; the Lovell telescope participated only on Oct 23. Baseline lengths ranged from 40k$\lambda$ to 3850k$\lambda$ with a uv hole of 200k-400k$\lambda$. A standard spectral setting was employed, in which the 4 spectral windows with %the 
128~MHz bandwidth and 1~MHz resolution were assigned with center frequencies at 4.880~GHz, 5.008~GHz, 5.135~GHz, and 5.264~GHz. These observations were calibrated using observatory provided scripts of CASA. Self-calibration and imaging were done using CASA. The Briggs weighting with the robust parameter of 0.5 was employed. In order to exclude contributions from extended emission (see Section 3.3), we use visibility data only with longer baseline lengths than 200k$\lambda$. The synthesized beam was 120~mas$\times$30~mas with the position angle of 20 degrees. The pixel size and the field of view were 15~mas and 4\arcmin, respectively. The rms noise level of the self-calibrated image was 30$\mu$Jy while the noise levels were significantly higher around strong point sources due to their sidelobes.  
{SrcI was not detected in this image.  }

\medskip
{\bf \section{DISCUSSION}}
{\bf \subsection{Disk Structure}}

\citet{Ginsburg2018} fitted the observed structure of the disk from B3 and B6 ALMA continuum observations at 50 and 20 mas resolution, respectively. They determined that the disk has a length of $\sim$100 AU, and vertical FWHM height of $\sim$20 AU.  
The integrated flux density for SrcI from 43 to 340~GHz has a spectral index of $\sim$2, consistent with optically thick dust emission \citep{Plambeck2016}.  However, images of the spectral index across the source reveal
an extensive region with spectral index $<$ 2 along the minor axis of the disk \citep{Wright2020,Wright2022}. 
Although the high opacity at 220 - 340 GHz hides the internal structure of the disk, images at 43 to 99 GHz reveal structure within the disk \citep{Wright2022}.

Table~2 summarizes the results of 
Gaussian fits to SrcI for the 6 to 340~GHz images. 
The disk major axis, at PA $\sim$ $-38$\degr, increases with observing frequency from 43 to 340~GHz which is expected as the dust optical depth increases.
At 6 and 14~GHz, however, the emission is extended at PA $\sim$55\degr, along the disk's minor axis.
The major axis of SrcI at centimeter wavelengths is nearly perpendicular 
to the major axis at millimeter wavelengths.  This is also evident in 
 \citet{Rodgriguez2017}; see their Table 2.
The 6 and 14 GHz Gaussian fits are overlaid on a 99~GHz image in Figure~\ref{fig:6GHz_image}.
Since the observations were obtained at different epochs (Table~1) we must correct the positions for the proper motion of
SrcI, moving at Dx=6.3, Dy=-4.2 mas~yr$^{-1}$ \citep{Goddi2011b}. 
We adjusted the 6~GHz and 14~GHz images of SrcI for proper motion to the 99~GHz epoch.
After this correction, the 6~GHz peak coincides with the center of the disk at 43 and 99 GHz within 5 mas, but the 14 GHz peak is offset 
to the SW by $\sim$70 mas.
The absolute position accuracy is estimated to be 20-30~mas (see the discussion in \citet{Forbrich2016} and in the VLA manuals\footnote{
https://science.nrao.edu/facilities/vla/docs/manuals/\\oss/performance/positional-accuracy}), so this
positional discrepancy suggests the possible presence of unresolved asymmetric structures that are not well fitted by a single Gaussian, or of time variability in the emission.

\begin{figure*}
% trim left bottom right top
\includegraphics[width=2.0\columnwidth, clip, trim=.5cm 4.5cm 0.5cm 1.2cm]{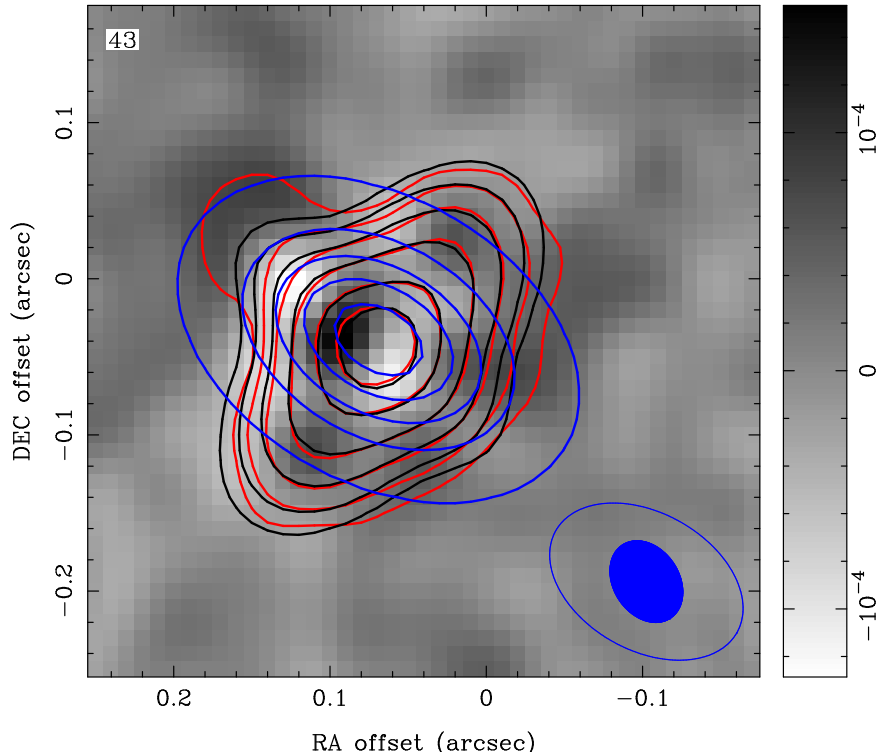}
\caption{43 GHz continuum emission and residuals to a 2-Gaussian model fit.
 The red contours show the 43 GHz continuum
emission convolved to 56 $\times$ 42 mas FWHM resolution. The black contours show the 2-Gaussian model, both with contours at 25, 50, 100, 200, 400, 600, 800, 1000 K . The pixel image shows residuals to the 2-Gaussian model fit (units Jy/beam).
The blue contours show the single Gaussian fit to the 6 GHz continuum emission. Contours at 100, 500, 900, 1300, 1700 K.
 The 43 GHz synthesized beam FWHM is indicated by the blue filled ellipse in the lower right. The 6 GHz Gaussian fitted FWHM is indicated by the blue open ellipse in the lower right.
 \label{fig:43GHz_2gauss_model+residual}}
\end{figure*}

% FIGURE 3
 \begin{figure*}
% trim left bottom right top
\includegraphics[width=2.0\columnwidth, clip, trim=.5cm 4.5cm 0.0cm 1.2cm]{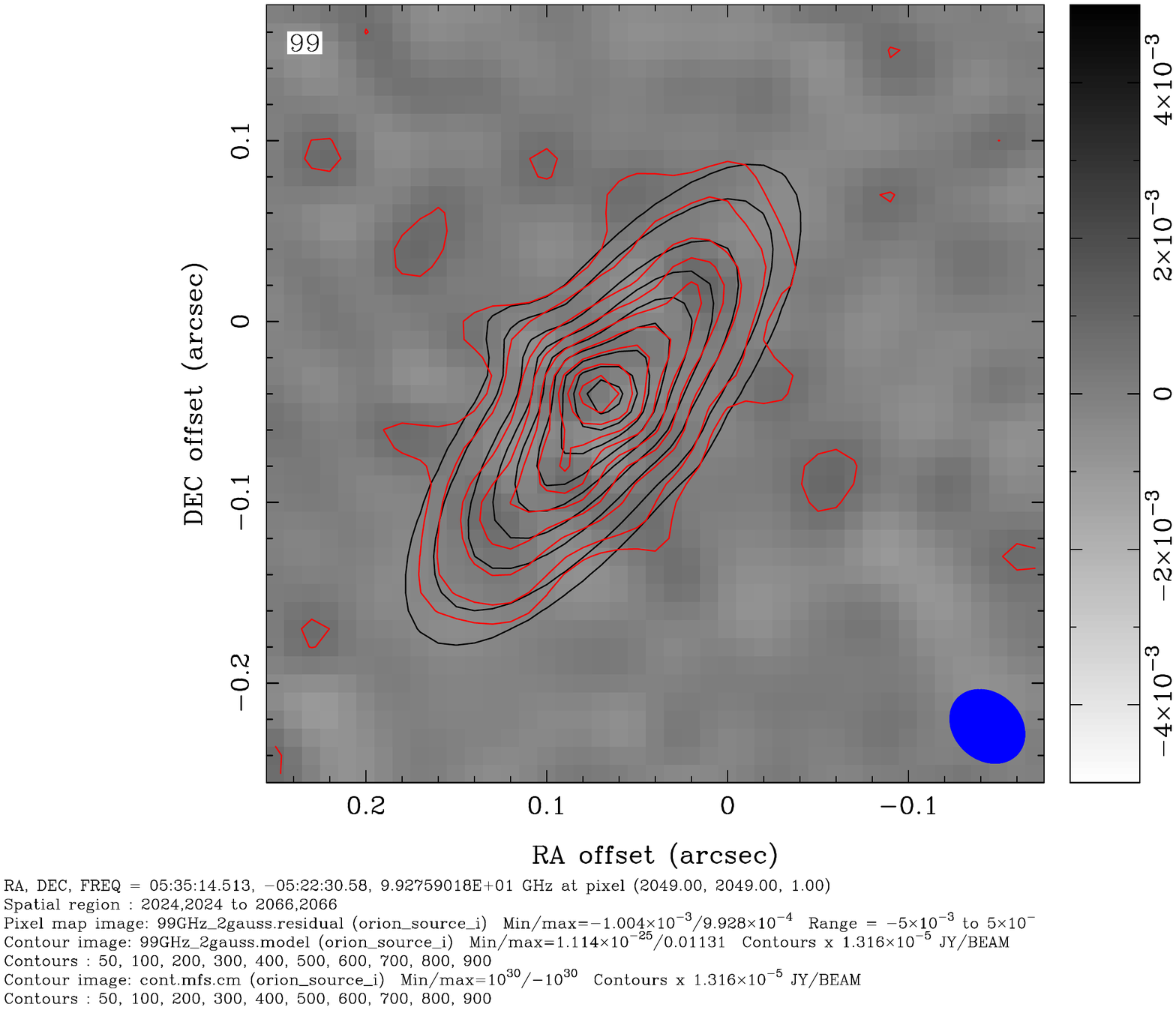}
\caption{99 GHz continuum emission and residuals to a 2-Gaussian model fit.
 The red contours show the 99 GHz continuum
emission at 56 $\times$ 42 mas resolution. The black contours show the 2-Gaussian model, both with contours at 50, 100, 200, 300, 400, 500, 600, 700, 800, 900 K . The pixel image shows residuals to the 2-Gaussian model fit (units Jy/beam). The synthesized beam FWHM is shown in the lower right.
 \label{fig:99GHz_2gauss_model+residual}}
\end{figure*}

\begin{figure*}
% trim left bottom right top
\includegraphics[width=2.0\columnwidth, clip, trim=.5cm 0.5cm 0.5cm 1.2cm]{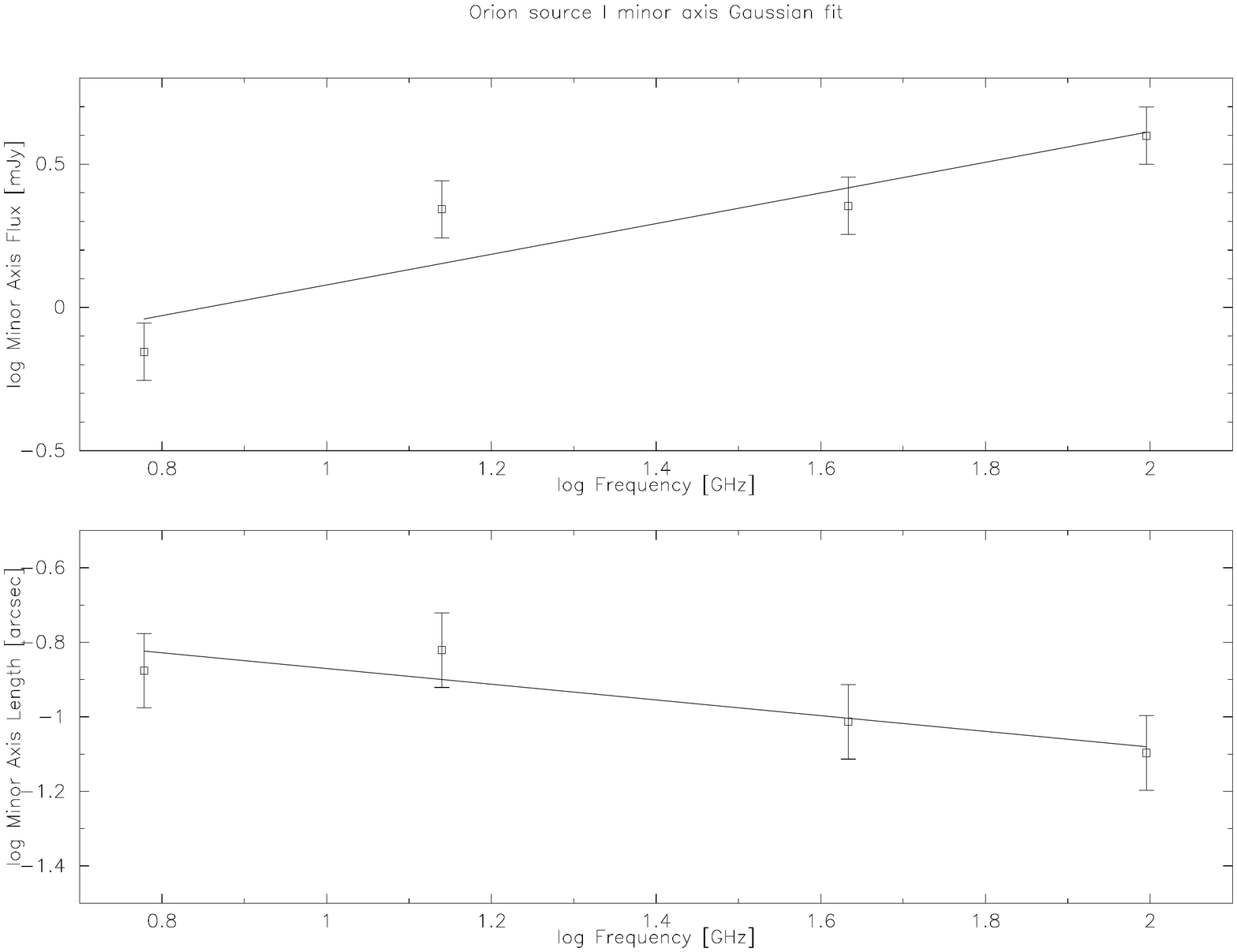}
\caption{Flux density and length (FWHM) of Gaussian fits along the minor axis of the 6, 14, 43, and 99 GHz continuum emission. Least squares fits are indicated by the straight black lines.
 \label{fig:minor_axis_fits}}
\end{figure*}

Weak emission extending along the minor axis of the disk also is apparent at 43 and 99~GHz.  We used 2-component Gaussian fits to estimate the flux and length of this structure, using the single Gaussian fit at 99 GHz along the disk major axis and the Gaussian fit at 6 GHz along the minor axis as initial parameters for the
fits.
Figures~\ref{fig:43GHz_2gauss_model+residual} 
and \ref{fig:99GHz_2gauss_model+residual} show the 43 and 99~GHz continuum images and model fits.  The parameters for the 2-Gaussian fits are included in Table 2.
The Gaussian components along the disk major axis are consistent with the single Gaussian fits.
The second Gaussian component is 
aligned with the disk minor axis, but the flux and length vary with frequency, and are plotted in Figure~\ref{fig:minor_axis_fits}. 
Note that the component along the disk minor axis is $\sim$23\% of the total flux at 43 GHz, and only $\sim$ 7\% at 99 GHz.

We also fitted the lower signal-to-noise 86 GHz continuum image, obtained
with smaller continuum bandwidth \citep{Wright2022}, with 2 Gaussian components. The Gaussian component
along the disk major axis is consistent with the single Gaussian fit. The flux of the
second Gaussian component is consistent with the 99~GHz result, but the position angle and size of 
this component depend strongly on the choice of image resolution fitted and have large estimated errors, 
hence are omitted from Table 2 and Figure~\ref{fig:minor_axis_fits}.

The 43 GHz images of SrcI in \citet{Goddi2011b} also suggest that there is a component along the minor axis of the disk. Their Figure 2 shows four 43 GHz continuum images of SrcI 
observed with the JVLA in the A configuration between November 2000 and January 2009.   
There appears to be no significant change in the morphology or P.A. of the structure over eight years.
The morphology is difficult to constrain in these data with different resolutions and sensitivity, but the flux could have changed by up to 20\%, and a 10\% change in structure is possible.  
We fitted a 2-component Gaussian model to the highest signal to noise image (AC952, 2009-01-12; Table 1).  A robust fit was obtained, consistent with the current 43 GHz image with a second Gaussian component along the disk minor axis with fitted parameters within 10\%.  

% -----------------------------------------------------------------------------------------------%

\medskip
{\bf \subsubsection {Compact hot spot.}}

This section discusses the small scale structure in the SrcI disk.
The Gaussian models in Table 2 are robust fits to the overall structure of SrcI, but do not preclude more complex small scale structure.
The high resolution 43 GHz observations of \citet{Reid2007}
 resolved SrcI into a disk plus a 
 compact, 2.2 mJy source at 34 mas resolution \citet{Reid2007} - Figure 1.
 \citet{Reid2007} also note that the disk appears to be warped with a component in PA 45\degr ~along the minor axis of the disk, which could indicate that there is a weak jet perpendicular to the disk.
\citet{Wright2022} - Figure 4 shows the warped ridge structure
in the 43, 86, and 99 GHz images at 30 mas resolution.
At 99 GHz there is an unresolved component with a position,  Right Ascension: 05:35:14.518, 
Declination: -05:22:30.608 and a peak brightness 4.4 +/- 0.2 mJy/beam at 30 mas resolution.
 The 43 GHz residuals to a 2-Gaussian fit (Figure~\ref{fig:43GHz_2gauss_model+residual})reveal complex small scale structure with peak residuals  +/- 5 sigma.
Fitting a point source to the peak residual at 43 GHz, gives a flux density 0.13 +/-  0.05 mJy in a 60 $\times$ 40 mas beam,
at Right Ascension: 05:35:14.519, Declination:  -05:22:30.611, which agrees with the 99 GHz position within a few mas. We note that our 43 GHz observations were self calibrated using SiO masers, and then aligned with the 99 GHz image, using the MIRIAD task, IMDIFF, which finds optimum parameters in a maximum likelihood sense, to minimize the difference between 43 and 99 GHz images, so we are actually
measuring the position of the peak relative to overall structure of the disk at 43 GHz.
\citet{Ginsburg2018}
detected a compact source at 224 and 93 GHz with flux $\sim$15 and 5.7 mJy respectively \citep{Ginsburg2018} - Table 4. The spectral index (SI) from 224 to 93 GHz $\sim$1.1. Including the 43 GHz point source fit at similar resolution \citep{Reid2007}, the SI from 224 to 43 GHz $\sim$1.2. The compact source could be evidence for a compact outflow from one of the binary protostars, or merger remnant.
The source is offset from the center of the disk, and appears to be extended along the disk major axis at 224 GHz.
As discussed by \citet{Ginsburg2018}, 
it is likely that the source is thermal but is more attenuated by the disk at 224 GHz than 93 and 43 GHz.  
If SrcI’s central source is a binary, the offset from the disk midpoint may indicate the
proximity of one of the protostars close to
 the inner edge of a circumbinary accretion disk. This suggests a binary orbit $\sim$5 au.
The orbital time is only $\sim$3 years. It will be interesting to see if the hot spot is variable on such a timescale.

  Higher resolution observations are required to see if the minor axis emission at 43 and 99 GHz is associated with the compact hot spot in the disk, rather than being centered on the 
  the circumbinary disk.
 More compact Gaussian or point source
fits to the 6 and 14 GHz images have higher
residuals; these data are not consistent with
a compact jet, however the data do not preclude a more complex extended source with a weak compact component.
The extended emission at 6 and 14~GHz favors a disk wind hypothesis, in which both the molecular and ionized outflow are  magneto-centrifugally driven 
   from different regions in a rotationally supported disk \citep{Blandford1982, Matsushita2018}. \citet{Vaidya2013}
modelled an MHD disc wind from a high-mass protobinary; specifically taylored on Orion source I.
   
High resolution observations at centimeter wavelengths are required to determine if SrcI harbors an ionized jet.
Observations of Orion-KL obtained with the e-MERLIN telescope at 5 GHz (6 cm wavelength) in 2021, provided a synthesized beam FWHM {120}$\times$30 mas with 40\% sidelobes which resolve out large scale structure.
SrcI was not detected in images, both with, or excluding short spacings (uv$<$200 klambda) to filter out large scale structure.
The upper limit on the SrcI brightness is 0.03 mJy/beam or 400 K (1-sigma). The intensity maximum around the position of Source I is about 0.1 mJy/beam or $\sim$ 1200 K at 3-sigma levels, which is consistent with the brightness temperature of the 43 GHz image.
The non-detection of SrcI by e-MERLIN is not consistent with the Gaussian fit to the images
from the JVLA in 2012-09-30 at 6 GHz with a synthesized beam 300$\times$194 mas.
Ionospheric wedges could spoil the e-MERLIN phase calibration, but the $\lambda^2$ dependence makes this unlikely at 5 GHz,
and self-calibration on strong compact sources in
the field of view around SrcI did not detect SrcI.
Consistent results were obtained on the three e-MERLIN
observation days. Twenty compact sources were detected in the field of view around SrcI at positions consistent with those measured by \citet{Forbrich2016}.
Some of these sources are resolved and attenuated in the e-MERLIN observations.
The 5-6 GHz structure of SrcI may be resolved by the high sidelobe structure of the e-MERLIN synthesized beam, and/or be time variable.

If the SrcI structure was significantly different from the Gaussian fits, it could have been resolved below the detection level in the e-MERLIN observations. For example,  
if the SrcI structure has
a north-south extent $\sim$0.3 arcsec it is attenuated by a factor 10 by the 40\% sidelobes in the e-MERLIN beam.  
High resolution with good uv-coverage (e.g. ngVLA or SKA) will be needed to map the SrcI  structure at centimeter wavelengths.

\begin{figure*}
% trim left bottom right top
\includegraphics[width=2.0\columnwidth, clip, trim=.5cm 1.5cm 0.5cm 0.2cm]{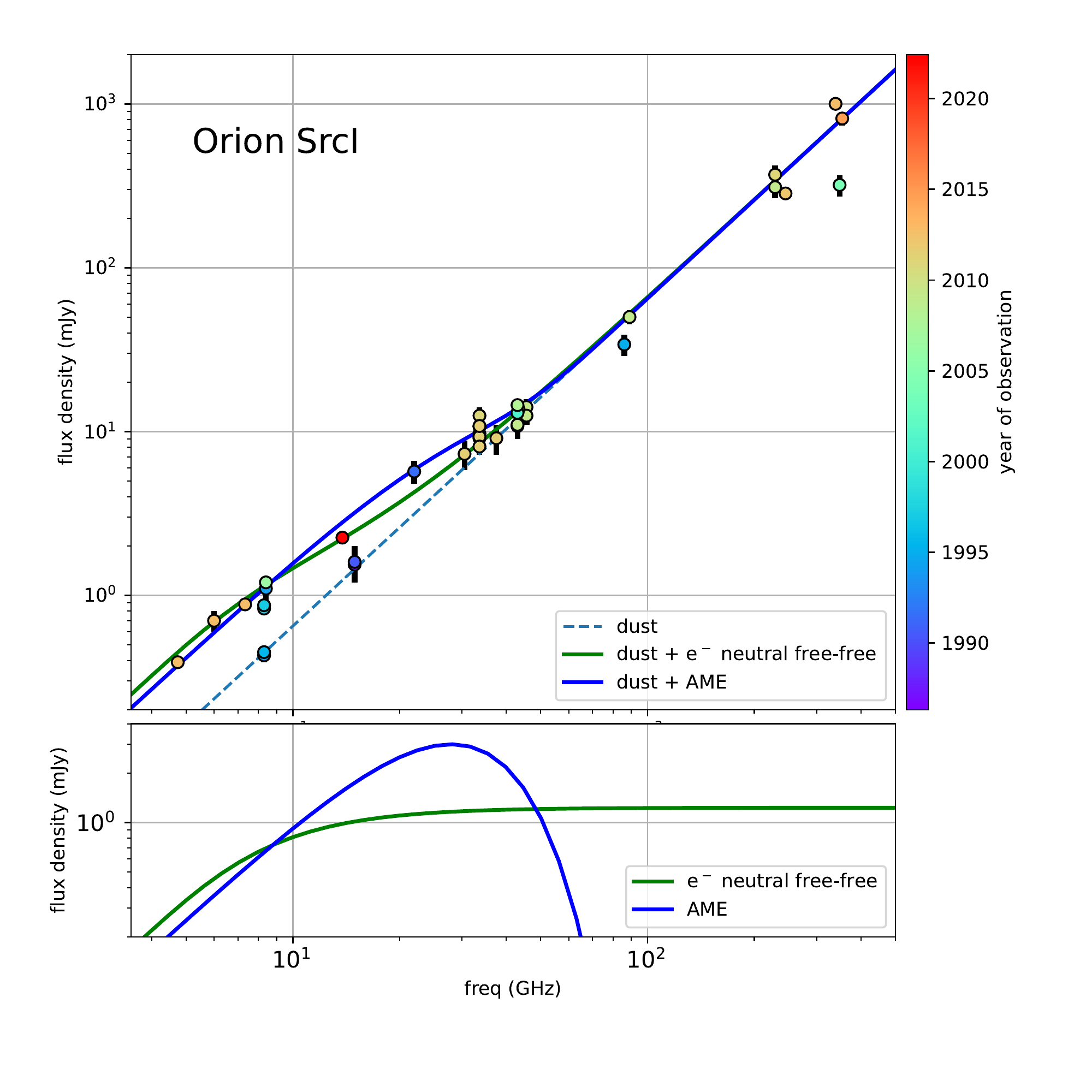}
\caption{ Spectral energy distribution of SrcI from cm to sub-mm wavelengths. References are given in \citep{Plambeck2016}. From $\sim$43 to 350 GHz the data follow a $\nu^2$ curve (dashed green line) that is fit to a flux density of 330 mJy at 229 GHz. At lower frequencies H-minus free–free emission (green curve),
and AME from spinning dust (blue curve) may become important. The observations are color-coded by year of observation.  There is some evidence that flux densities have increased slightly since the 1990s. The published error bars  for most of the flux measurements are smaller than the plot symbols.   
 \label{fig:SrcI_SED+AME}}
\end{figure*}

Recent radio continuum surveys have detected a number of high-mass YSOs associated with radio jets \citep{Purser2016, Sanna2018}. They are thought to be driven from the closer vicinity to the central YSOs given their higher velocities, of a few 100 km/s, than molecular outflows. In the
MHD disk wind model, a higher-velocity outflow is predicted to be launched in the rotating disk with a larger rotation velocity than the molecular outflow \citep{Matsushita2018}.
 \citet{Vaidya2013}
 found a lower MHD wind velocity for the protobinary disk model for Orion SrcI.
\medskip
{\bf \subsection{SED}}
The spectral energy distribution of SrcI from 4-340~GHz is shown in Figure~\ref{fig:SrcI_SED+AME}.
To search for evidence of time-variability in the SrcI fluxes, we have color coded 
the data by year of observation.  It does appear possible that SrcI has brightened
somewhat over the 30 year period spanned by these measurements, although
much of the scatter in the data probably is due to the difficulty in measuring the
flux of a compact source that is embedded in a complex, lumpy, extended medium.

At frequencies $>$43~GHz, the SED appears to follow a $\nu^2$ curve, consistent with optically
thick emission from dust in the SrcI disk.  At lower frequencies, however, most of the flux 
densities -- including the 6 and 14~GHz measurements reported in this paper -- lie above this curve.  Our images at 6~ and 14~GHz show an extended region with SI $<$ 2 along the minor axis. \citet{Wright2022} discuss mechanisms which could produce an SI $<$ 2.  Emission along the minor axis of
the disk could arise from a separate emission mechanism.  We consider several
possible mechanisms in the sections below.  Example curves for two of these,
electron-neutral free-free emission, and dipole radiation from spinning dust 
(Anomalous Microwave Emission) are shown in the lower panel of Figure~\ref{fig:SrcI_SED+AME}.

\medskip

{\bf \subsubsection{Thermal dust emission}}
We first consider the possibility that the ``excess'' flux at low frequencies is 
thermal emission from hot dust close to the central object.  This dust would emit at
all frequencies, of course, but at mm wavelengths the emission might be absorbed
by an extended envelope of cooler foreground dust.  The
Gaussian fit to the 6~GHz image indicates that the measured flux density of 0.7~mJy
originates from an 0.14$\times$0.09$\arcsec$ region.  This implies a 6~GHz brightness 
temperature of 3000~K, higher than the sublimation temperature of dust. However,
radiation pressure from the protostars may shock and expel dusty material along the minor axis of the disk \citep{Hosokawa2009}.
The wide angle outflow observed in SiO emission shows that material is being ablated from the disk.  Emission at cm wavelengths would require a population of pebble-sized grains 
in this outflow, which seems unlikely.

% -----------------------------------------------------------------------------------------------%

\medskip
{\bf \subsubsection {Ionized outflow.}}

An ionized outflow could have a spectral index $<$ 2.  The bulge to the NE along the minor axis at 43 GHz, and a smaller one at 99 GHz have a spectral index $\sim$1.
The Gaussian fits at 6, 14, 43 and 99~GHz  show a component along the disk minor axis whose flux and length vary with frequency consistent with an ionized outflow from 6 to 99~GHz.
Figure~\ref{fig:minor_axis_fits} plots the flux density and length (FWHM) of Gaussian fits along the minor axis of the 6, 14, 43 and 99~GHz continuum emission. 

The total flux in the minor axis component scales with frequency as $\nu^{0.5 +/- 0.1}$ from 6 to 99~GHz.
The spectral index between 6 and 14 GHz steepens to ${1.35 +/- 0.1}$.
At 6 GHz, \citet{Forbrich2016} derived a spectral index 1.86 +/- 0.26 from the two 1 GHz basebands centered at 4.736 and 7.336 GHz.
The steeper spectral index at lower frequency is consistent with the free-free emission fitted to the integrated flux measurements shown in \citet{Plambeck2016} - Figure 14.
The length of the minor axis component decreases with frequency as $\nu^{-0.2 +/- 0.1}$ between 6 and 99~GHz.
 This is consistent with the ionized outflows discussed by \citet{Reynolds1986} who shows that a collimated ionized outflow can exhibit a wide range of parameters with spectral index from -0.1 to 2,
varying along the outflow depending on the velocity gradients
and confinement (see \citet{Reynolds1986}, Figure 2),
and notes that variations in the flow along the outflow may yield observable spectral features. The residuals
in Figure~\ref{fig:43GHz_2gauss_model+residual} presented here, may represent such spectral variations along the outflow.
\medskip
{\bf \subsubsection {Free-free emission.}} 
The best evidence for free-free emission would be the detection of hydrogen
recombination lines.  These radio recombination lines (RRLs) will be visible only at frequencies greater than the turnover frequency where the continuum emission becomes optically thin.
We have only upper limits on recombination lines.
The ratio of the radio recombination line (RRL) to continuum intensity expected from pure free-free emission is $\sim$1-2 at cm wavelengths \citep{Gordon2002}, so an upper limit on the RRLs would directly limit the free-free contribution to the continuum where any free-free component is optically thin. The line/continuum ratio increases with frequency so we looked for recombination lines at mm wavelengths.
Assuming the outflow is co-rotating with the
molecular outflow observed in SiO, we estimate a FWHM linewidth $\sim$30 km/sec.
According to eqn 14.29 in \citet{Wilson2009}, we find a line/continuum ratio of about 4 at 230 GHz.
\citet{Plambeck2016} analysed the continuum flux densities measured for SrcI from 4 to 690 GHz.
Free–free emission may become important where the dust emission becomes optically thin.  \citet{Plambeck2016} - Figure 14 show fits for a spectral index of 2, with excess emission at submillimeter wavelengths from dust emission, and
free-free emission which fits the excess flux at cm wavelengths.
 The H26$\alpha$ (353.6 GHz) and H21$\alpha$ (662.4 GHz) hydrogen recombination lines were not detected. \citet{Plambeck2016} used a velocity range 5 $<$ v$_{lsr}$ $<$ 23 km s$^{-1}$  to establish upper limits on the intensities of the recombination lines.
 No recombination lines were detected in the SrcI disk with an upper limit 5\% of the total continuum flux integrated in a 0.2$\times$0.2$''$ box centered on SrcI.
 We re-examined our observations of the H30$\alpha$ line
 at 231.9 GHz. We made spectral line images with 65$\times$35 mas and 5 km/s resolution.  The H30$\alpha$ recombination line was not detected in a velocity range $-$50 to +50 km/s.
The peak brightness (+/$-$ 3 sigma) was 6 K , only 4 \% of the disk continuum brightness 440 K. 
The ionization may be very different where the disk is exposed to radiation from the protostars, or shocks generated in the outflow \citep{Shull1979}, so an overall ionization fraction may not represent conditions in different parts of the disk. These observations do not exclude ionized emission at the 4\% level in the disk, and 25\% in the lower brightness bulge to the NW, observed at 43 GHz. 
 There may be substantial extinction by cooler dust that would obscure recombination lines, even along the minor axis of the disk.
\citet{Baez-Rubio2018} modelled SrcI as dust and H-minus emission (see below). Their model in table 3 \& 4 show that dust opacity reduces the RRL detectability, rendering this diagnostic for ionization fraction less definitive.

\citet{Plambeck2016} -Figure 14 plot a fit to the low frequency integrated fluxes with a free-free component with n${_e}$ = 5$\times$10$^{-6}$ cm$^{-3}$. 
Using the fitted size of the observed emission at 6 to 43 GHz, we obtain a mass $\sim$8$\times$10$^{-8}$ $M_{\odot}$ for electron-proton free-free emission.

The outflow in Figure~\ref{fig:99GHz+43GHz+h2o+AlO_N=6-5_B6+3mmSiOmasers}  appears to expand and dissipate close to where the SiO J=1-0 and 2-1 ouflows mushroom out from a co-rotating SiO column (see \citet{Hirota2020} - Fig 2).
Since the SiO outflow is rotating and expanding, an ionized outflow may also be rotating, and would expand and dissipate where the rotation velocity is similar to velocity of SrcI
through the ambient gas. Our H30$\alpha$ observations with 5 km/s resolution  over 100 km/s would detect a rotating outflow, but a larger line width may be below the detection threshold.
The RRL lines may have very broad line-widths as for example in Cep A IRS1 \citep{Jimenez-Serra2011}.
 If the jet is highly turbulent or has a large opening angle the line width could be a few hundred km/s.
Using Eq. (24) \citep{Anglada2018}, the line-to-continuum ratio is $\sim$0.01 for a line width $\sim$200 km/s, and would not be detected in these observations.

\medskip
{\bf \subsubsection {Electron–Neutral Free–Free Emission}}
 Electron–Neutral Free–Free Emission
\citep{Reid2007, Plambeck2016} may be responsible for the higher flux densities in the SrcI disk below 43 GHz. 
Electron–Neutral Free–Free (H-minus) emission occurs for temperatures
between 1000 and 4500 K, where Na and K would be collisionally ionized but H and H$_2$ are neutral, may be relevant in an ionized outflow from SrcI.
\citet{Hirota2015} derived a spectral index 1.60 +/- 0.24 for a combination of dust and H-minus emission to explain the  emission from the SrcI disk for a range of total H densities from 10$^{11}$ to 10$^{14}$ cm$^{-3}$, and temperatures 1200 to 3000 K. 
\citet{Baez-Rubio2018} modelled SrcI as dust and H-minus emission, assuming a cylindrical geometry for a neutral disk with dust, and derived a H density 0.5 to 7 $\times$ 10$^{11}$ cm$^{-3}$, with temperatures $\sim$ 1700 K at a radius 20 AU 
estimated from 43 GHz images.
Higher resolution and higher frequency data suggest optically thick dust as the opacity source in the SrcI disk \citep{Plambeck2016, Hirota2016a}.
Although H-minus emission does not fit the overall spectral index of the disk where the emission is dominated by optically thick dust emission above $\sim$100 GHz, it might explain the observed SI for the emission along the minor axis, and in the outflow where our observations from 6 to 99 GHz show a component extended along the minor axis with an SI $<$2.

We calculated the H-minus free-free emission expected from gas in a 56$\times$36$\times$36 AU rectangular prism (0.14$\times$0.09$\times$0.09 arcsec) for a hydrogen density 4$\times$10$^{10}$ (n[H2] = 2$\times$10$^{10}$ cm$^{-3}$), and temperatures of 1500-1900 K. 
This is roughly the deconvolved FWHM sizes at 6 and 14 GHz, and would fit inside the boundary defined by SiO v=1 and v=0 masers. The total mass of gas in this prism is 0.008 $M_{\odot}$. If this material is flowing out from SrcI at 18 km/sec, this implies a mass outflow rate of 10$^{-3}$ $M_{\odot}$/yr.
The density and Te are similar to those derived by \citet{Hirota2015} and \citet{Baez-Rubio2018} (which assume different geometries), but an outflow 10$^{-3}$ $M_{\odot}$/yr is $\sim$ 100$\times$ that estimated from SiO, and would deplete the disk mass unless the accretion rate is also high \citep{Wright2022}.
Figure~\ref{fig:SrcI_SED+AME} shows the excess low frequency emission that could result from H-minus free-free emission in the total flux SED for SrcI. In this simple model, N$_e$ $\sim$10$^4$ cm$^{-3}$, and the fractional ionization $\sim$10$^{-7}$ which may be too low to support an MHD outflow. Of course, the ionization might be higher where the disk is exposed to the protostars and shocks.

For regular free-free emission, the radiation arises from ``collision'' between H${^+}$ (protons) and electrons, and each such collision provides an opportunity for the H${^+}$ and electron to recombine.  By contrast, for H-minus free-free emission, the radiation originates from collisions between neutral H and electrons.
 The exceedingly rare collisions between Na${^+}$ and e${^-}$ can result in recombination. 
The salt line at 232.51 GHz which traces the rotation of the disk, and correlates with the dust emission at the ends of the disk \citep{Ginsburg2019}, has emission at v$_{lsr}$ = 5 km s$^{-1}$ along the minor axis where it is exposed to ionizing radiation from the center of the disk.
Na${^+}$ radio recombination lines, but not hydrogen recombination lines, would strongly suggest H-minus emission.
As discussed by \citet{Ginsburg2019}, the lowest excited electronic energy
levels for NaCl is about 5 eV above the
ground state,  close to the photodissociation threshold
(Zeiri \& Balint-Kurti 1983; Silver
et al. 1986a), which may be dissociated by UV photons. UV emission may be produced in
shocks in the outflow, and $\sim$ 5 eV photons could be produced by the central $\sim$ 4000 K photosphere \citep{Testi2010}, and ionize NaCl along the minor axis of the disk where the opacity is low.
\citet{Baez-Rubio2018} models predict the expected radio recombination line intensities in Na26$\alpha$ and K26$\alpha$ (see their tables 3 and 4), which are below their observation RMS limits of $\sim$49 and 20 mJy respectively.
 
\medskip 
 {\bf \subsection{Alternatives to an ionised outflow}}
 
 The lower spectral index along the minor axis of the SrcI disk, and the bulge to the NW at 43 GHz might also be explained by different conditions along the rotation axis of the disk. The wide angle outflow observed in SiO emission shows that material is being ablated from the disk.
 Radiation pressure from the protostars may shock and expel dusty material along the minor axis of the disk \citep{Hosokawa2009}. 
 We estimated radiation pressure  at radius R, where T =1700 and 500 K as nominal inner and outer disk surface defined by the observations.
For a spherical geometry, and a central luminocity 10$^4$ $L_{\odot}$, we obtain inner and outer radii R $\sim$5.5 AU and 63 AU.
A disk geometry with surface 2$\pi$R$^2$(1 + d/R) with d/R = 0.1 gives similar R values.
Equating the radiation and gas pressures, assuming that the gas has high opacity and dust is mixed with the gas, we estimate a velocity dispersion, v, for H$_2$,
and density for pressure balance.
For the inner surface  T $\sim$1700 K and  R $\sim$5 AU, v $\sim$4 km/s, and N[H2] $\sim$10$^{11}$ cm$^{-3}$.
For the outer surface, T $\sim$500 K and R $\sim$50 AU, v $\sim$2 km/s, and N[H2] $\sim$ 7$\times$10$^8$ cm$^{-3}$.
This suggests that densities less than the
above could be ablated and pushed away by radiation pressure.
The bulge observed at 43 GHz could be partially ionised material ablated from the disk. 
Higher temperatures would strip off the grain mantles, releasing H$_2$O, and other molecules with peak emission close to the disk \citep{Wright2020}.

\begin{figure*}
% trim left bottom right top
\includegraphics[width=2.0\columnwidth, clip, trim=.5cm 4.5cm 0.5cm 1.2cm]{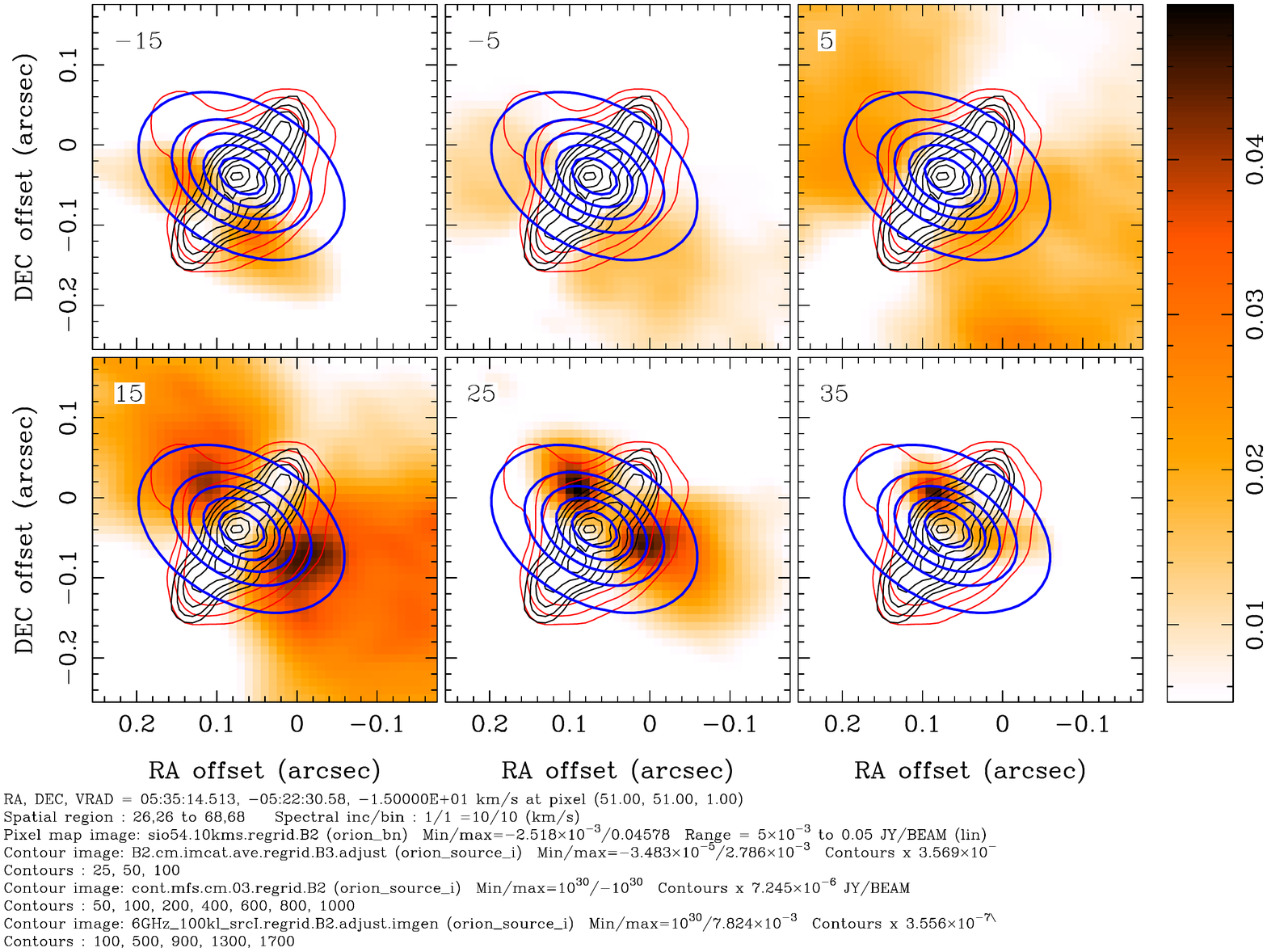}
\caption{SiO $v=0$ $J=5-4$  emission at 217.10498 GHz overlaid on contours of the 99, 43 and 6 GHz continuum emission. The thick black contours at 99 GHz at 50, 100, 200, 300, 400, 500, 600, 700, 800, 900, 1000 K with a 30 mas FWHM show the inner structure in the disk. The blue contours show the deconvolved Gaussian fit to the 6 GHz image. Contours at 100, 500, 900, 1300, 1700 K. The red contours at 43 GHz at 25, 50 and 100K are convolved to 56 $\times$ 42 mas FWHM resolution to show the structure at the edge of the disk. The color image shows SiO $v=0$, $J=5-4$ emission at 10 km s$^{-1}$ intervals (units Jy/beam). Beam FWHM 54 x 34 mas in PA 65 \degr.
 \label{fig:99+43+6GHz+sio54_10kms}}
\end{figure*}

\begin{figure*}
% trim left bottom right top
\includegraphics[width=2.0\columnwidth, clip, trim=.5cm 4.3cm 0.5cm  1.2cm]{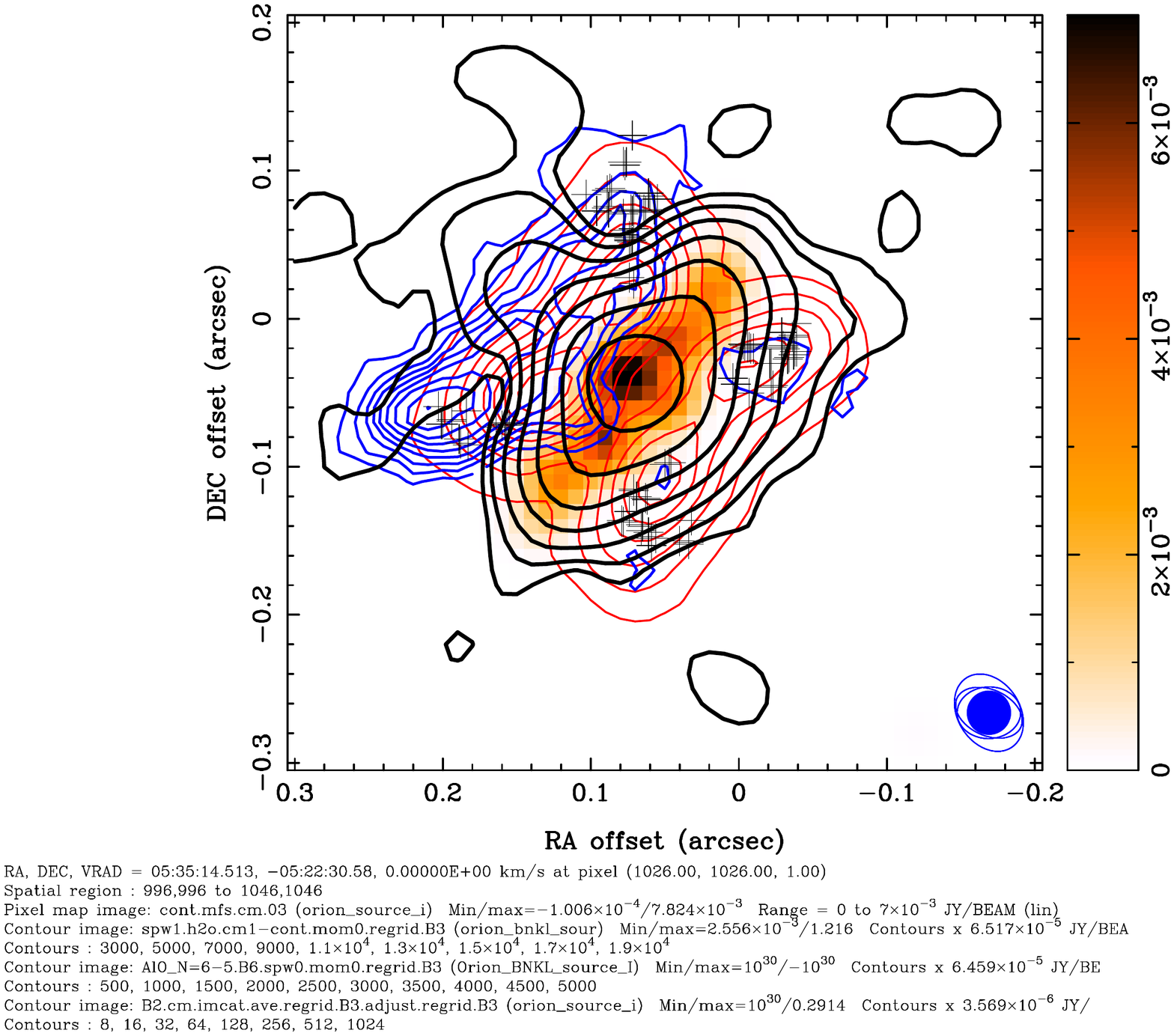}
\caption{ AlO, H${_2}$O, and SiO maser emission,  overlaid on contours of the 43 GHz, and 99 GHz continuum emission.
The black contours at 8, 16, 32, 64, 128, 256, and 512 K show the low brightness emission at the edge of the disk at 43 GHz.
The black crosses indicate the centroid positions of the  $v=1$, $J=2-1$ SiO masers mapped by \citet{Issaoun2017}. 
 Red contours show the \ce{H2O} emission at 232.6867 GHz integrated over a velocity range (-50 to 50) km/s. Lowest contour 3000, contour interval 2000 K km s$^{-1}$.
Blue contours show the AlO N=6-5 emission integrated over -50 to 50~\kms. Lowest contour 500, contour interval 500 K km s$^{-1}$.  The 30 mas convolving beam FWHM for the 99 GHz continuum color image (units Jy/beam), is indicated in blue filled circle. 
The convolving beam at 43 GHz, 56 $\times$ 42 mas FWHM resolution, is shown in the lower right. The FWHM beams for the \ce{H2O} emission of 0.05 $\times$ 0.03 arcsec in PA 65$^{\circ}$, and for the AlO N=6-5, 0.04 $\times$ 0.03 arcsec in PA -88$^{\circ}$, are also shown as the open ellipses in the lower right.
\label{fig:99GHz+43GHz+h2o+AlO_N=6-5_B6+3mmSiOmasers}}
\end{figure*}

The distribution of  SiO, H${_2}$O and AlO emission provide additional chemical tracers of the outflow from SrcI.
Figure~\ref{fig:99+43+6GHz+sio54_10kms} shows that the molecular outflow from the disk, mapped here in SiO $v=0$, $J=5-4$ emission at 5 km/s resolution, extends across the outflow seen in the 43 GHz continuum emission.

We mapped the \ce{H2O} emission at 232.687 GHz, AlO N=6-5 line at 229.69387 GHz, and 
the AlO N=9-8 line at 344.4537 GHz \citep{Wright2020}.
Figure~\ref{fig:99GHz+43GHz+h2o+AlO_N=6-5_B6+3mmSiOmasers} shows 229.694~GHz AlO emission overlaid on 43 GHz continuum and \ce{H2O} emission at 232.687 GHz. \ce{H2O}, AlO, and SiO maser emission, are
located along the sides of the outflow mapped at 43 GHz.

AlO emission peaks are further out than \ce{H2O}, and are much brighter on the NE side of SrcI where the outflow at 43 GHz is seen, and the hot spot in the disk is located.
 The distributions of AlO emission and \ce{H2O} shown here are consistent with those for AlO N=13-12 and N=17-16 emission lines at 497 and 650 GHz
\citep{Tachibana2019}, and \ce{H2O} at 463 GHz \citep{Hirota2017}. \citet{Tachibana2019} suggest that the AlO is produced at the base of the outflow, and condenses further out into the outflow.  
Our observations { with $\sim 4\times$ higher angular resolution,} show that AlO emission { peaks} downstream of the \ce{H2O}, and can be traced back to the ridge of continuum emission shown in 
Figure~\ref{fig:99GHz+43GHz+h2o+AlO_N=6-5_B6+3mmSiOmasers} at 99 GHz.  AlO may be produced by grain destruction and oxidation by O released by the dissociation of \ce{H2O}, or released as AlO further out in the outflow than the \ce{H2O} emission \citep{Wright2020}.
These observations strongly suggest that the outflow is closely associated with grain chemistry in the outflow.
\medskip
{\bf \subsubsection{Anomalous Microwave Emission}}

The excess emission below $\sim$100 GHz
could also be caused by anomalous microwave emission (AME) which would lower the SI along the minor axis and in the outflow. \citet{Rafikov2006} shows that AME from small spinning dust grains, at $\nu<$ 50 GHz can exceed thermal dust  emission by a factor of several if more than $\sim$5\% of the carbon is in nanoparticles. 
Large grains are shattered by grain collisions in shocks. \citet{Jones1996} show that for shock velocities 50 to 200 km/s, 5\% to 15\% of grains $>$ 0.5$\mu$m  can be shattered into $\sim$0.1$\mu$m fragments on time scales less than 10$^8$ years in the warm medium.
Nanograin production from larger grains is also predicted to occur close to sources of intense radiation by rotational disruption of dust grains by radiative torques \citep{Hoang2019,Hoang2021}, and in shocks \citep{Tram2019}.
 The formation of water ice mantles can affect the capacity of nanosilicates to generate AME \citep{Guiu2021}.
Destruction of dust grains seems likely in the outflow to release H2O, SiO and AlO. Although dust grain mantles may not survive in the $\sim$500+ ~K disk, fresh infalling material could bring in grains with mantles.
Figure~\ref{fig:SrcI_SED+AME} shows the excess low frequency emission that could result from AME from spinning dust grains in the total flux SED for SrcI.
The AME flux may be time variable depending on the excitation of the spinning dust. If the spinning dust emission depends on time variable shocks, and radiation, it too may be variable.

\medskip
{\bf \subsubsection{Non-thermal Emission}}
 The VLBA survey of the Orion region  at 7 GHz \citep{Forbrich_2021} did not detect SrcI at 4 epochs between 2015-10-26 and 2018-10-26.
 The synthesized beam FWHM ranged from 4.7$\times$1.6 mas, to 2.8$\times$1.2 mas with a shortest uv-spacing $\sim$3M$\lambda$ corresponding to a largest angular size $\sim$100 mas. The small synthesized beam and high brightness temperatures in the VLBA observations require non-thermal emission \citep{Forbrich_2021}.
 The VLBA observations set a 5.5-sigma upper limit of 0.21 mJy for SrcI structure smaller than $\sim$100 mas. For optically thick emission, this may be resolved out entirely, without any information on non-thermal emission.
\medskip 
 {\bf \subsubsection{Episodic Emission}}
 Many of the other sources in the FOV, are variable in these VLBA observations.
Variable emission at 5 - 7 GHz could be associated with episodic accretion events onto the embedded protostars with structural or excitation changes in the ionized or non-thermal emission. 

A possible reason for non-detection by e-MERLIN is
an increase in size from 0.1 to $\sim$0.5 arcsec in the $\sim$9 yr between the JVLA and e-MERLIN observations, which would correspond to a velocity $\sim$80 km/s in the plane of the sky.

Other massive young protostars have radio jets with high proper motions. In Cepheus A, the radio jet has a proper motion $\sim$500 km/s \citep{Curiel2006}.
\citet{Cunningham2009} proposed that the precessing jet is driven by the massive protostellar source HW2, with close passages of a companion in an eccentric orbit, which perturbs the disk and triggers accretion and episodes of jet production.

In the high mass proto-binary system, IRAS
16547–4247, the ionized jet appears to come from only
one of the binary protostars \citep{Tanaka2020}. Such a jet might be associated with episodic, or periodic accretion corresponding to the orbital period of the binary. 

Similar to these high-mass YSOs \citet{Purser2016,Sanna2018}, there could be a compact radio jet in SrcI emanating from the inner region.
Considering the large specific angular momentum carried by the molecular outflow, it is unlikely that a radio jet in SrcI could entrain the outflow traced by the $Si^{18}O$ line \citep{Hirota2017}. If higher resolution observations do not identify a radio jet driven from SrcI, this would also suggest that the molecular outflow in SrcI is not driven by an entrainment in a jet.

For Orion SrcI, with a binary separation 1 - 9 AU suggested by the BN/SrcI explosion \citep{Goddi2011}, the orbital period $\sim$0.3 to 10 yr. The outflow and ionization of the material along the minor axis in SrcI might be expected to echo accretion flows, leading to the intriguing possibility of mapping accretion events, and determining the orbital parameters of compact binary systems like Orion SrcI, from time variations in the centimeter wavelength emission.
\medskip
{\bf \section{SUMMARY}}

1. We present images at 6 and 14~GHz of emission from Source I in Orion-KL with $\sim0.3''\times0.2''$ resolution. The 6 and 14~GHz emission trace structure along the minor axis of the disk also apparent in images from 43 to 340~GHz.

2. Although the high opacity at 220 - 340 GHz hides the internal structure, images at 43 - 99~GHz reveal structure within the disk.
Images of the spectral index distributions show an extensive region with spectral index $<$ 2 along the minor axis.

3. Gaussian fits at 6, 14, 43 and 99 GHz  show a component along the disk minor axis whose flux and length vary with frequency consistent with an ionized outflow.

4. The 43 GHz images from 2002 to 2009 \citep{Goddi2011}, are consistent with this component along the minor axis of the disk within $\sim$10\%. There is no significant change in the morphology or P.A. of the structure at 43 GHz over eight years.

5. Source I was undetected in higher angular resolution e-MERLIN observations at 5 GHz in 2021.
The 5-6 GHz structure of SrcI may be resolved out by the high sidelobe structure of the e-MERLIN synthesized beam, or  be time variable.
If the SrcI structure was significantly different from
the Gaussian fit, it could have been resolved
below the detection level in the e-MERLIN observations.

6. The extended emission at 6 and 14~GHz favors disk wind hypothesis, in which both the molecular and ionized outflow are  magneto-centrifugally driven in a rotationally supported disk.

7. The correlation with SiO masers, H2O, and AlO, suggest that the 6 to 99 GHz
extension along the minor axis is associated with dust grain destruction, and a dusty, partially ionized outflow of material pushed out from the disk by the luminosity of the star(s).

8. Future high resolution and well calibrated observations at centimeter wavelengths will help to distinguish between the different emission mechanisms, and look for time variations in the structure of the disk and outflow.

% --- table of observations --- %
\begin{deluxetable*}{cccccc}
\tabletypesize{\small}
\tablecaption{Observations}
\tablecolumns{6}
\tablenum{1}
\tablehead{
\colhead{freq} &
\colhead{project code} &
\colhead{date} &
\colhead{time} &
\colhead{synth beam}  &
\colhead{baseline} \\
\colhead{(GHz)} & & & 
\colhead{(min)} &
\colhead{(milliarcsec)} &
\colhead{(meters)}
}
\startdata
5  & e-MERLIN/CY12202 & 2021-10-23,25,26 & 1023 & 120$\times$30 at PA 20 & 2400 - 217500 \\
6  &  VLA/SD630     &  2012-09-30  &  291  &   300$\times$194 at PA 30  & 500 - 36600  \\
14  &  VLA22A-022   &  2022-06-04  &  200  &   280$\times$160 at PA 42  & 500 - 36600  \\
 43  &  VLA/AC952     &  2009-01-12  &  198  &   58$\times$39 at PA 3  & 500 - 36600  \\
 43  &  VLA/18A-136     &  2018-03-06  &  291  &   39$\times$34 at PA 1  & 500 - 36600  \\
 86 and 99  &  2017.1.00497.S	&  2017-10-12  &  158  &   45$\times$36 at PA 47  & 40 - 16200 \\
224  &  2016.1.00165.S	&  2017-09-19  &   44  &   39$\times$19 at PA 66  & 40 - 10500 \\
340  &	2016.1.00165.S  &  2017-11-08  &   45  &   26$\times$11 at PA 58  & 90 - 12900 \\
\enddata
\end{deluxetable*}

% -------------------------------------- TABLE -----------------------------------------------
\begin{deluxetable*}{CCCCC}   % capitals indicate math mode
\tabletypesize{\small}
\tablecaption{Measured sizes and flux densities for source I}
\tablecolumns{5}
\tablenum{2}
\tablehead{
\colhead{freq} & 
\colhead {Gaussian} &
\colhead{deconvolved size} &
\colhead {PA} &
\colhead{integrated flux} \\
\colhead{(GHz)} &
\colhead {components} &
\colhead{(arcsec)} &
\colhead{(deg)} &
\colhead{(mJy)}
}
\startdata
6 & 1 &
0.133 \pm 0.01 \times 0.088 \pm 0.010 & 61\degr \pm 3\degr & 0.7 \pm .1 \\
14 & 1 &
0.151 \pm 0.01 \times 0.103 \pm 0.010 & 53\degr \pm 4\degr & 2.2 \pm .2 \\
43 & 1 & 0.099 \pm 0.002 \times 0.057 \pm 0.002 & -40.4\degr \pm 3\degr & 10 \pm 1 \\
43 & 2 & 0.125 \pm 0.002 \times 0.044 \pm 0.002 & -39.4\degr \pm 1\degr & 7.6 \pm 1 \\
43 & 2 & 0.097 \pm 0.002 \times 0.030 \pm 0.002 & 53\degr \pm 3 \degr   & 2.3 \pm 0.5 \\
86 & 1 &
0.142 \pm 0.005 \times 0.044 \pm 0.001 & -38.0\degr \pm 1.0\degr & 48 \pm 5 \\
99 & 1 & 0.151 \pm 0.005 \times 0.044 \pm 0.002 & -37.8\degr \pm 1.3\degr & 58 \pm 6 \\
99 & 2 & 0.163 \pm 0.005 \times 0.041 \pm 0.002 & -37.8\degr \pm 1.0\degr & 55 \pm 6 \\
99 & 2 & 0.081 \pm 0.005 \times 0.029 \pm 0.002 & 56\degr \pm 3\degr      & 4 \pm 1 \\
224 & 1 &
0.197 \pm 0.003 \times 0.042 \pm 0.003 & -37.3\degr \pm 0.4\degr & 256 \pm 25\\
340 & 1 &
0.234 \pm 0.005 \times 0.042 \pm 0.002 & -37.4\degr \pm 0.3\degr & 630 \pm 63 \\
\enddata
\label{table2}
\end{deluxetable*}

\medskip
\acknowledgments

 We thank the referee for a careful reading and good suggestions which have improved the presentation of this paper.
This paper makes use of ALMA data listed in table 1.

ADS/JAO.ALMA\#2016.1.00165.S,
ADS/JAO.ALMA\#2017.1.00497.S. 
 ALMA is a partnership of ESO (representing its member states), NSF (USA) and NINS (Japan), together with NRC (Canada), MOST and ASIAA (Taiwan), and KASI (Republic of Korea), in cooperation with the Republic of Chile. The Joint ALMA Observatory is operated by ESO, AUI/NRAO and NAOJ. e-MERLIN is a National Facility operated by the University of Manchester at Jodrell Bank Observatory on behalf of STFC.
The National Radio Astronomy Observatory is a facility of the National Science Foundation operated under cooperative agreement by Associated Universities, Inc.
TH is financially supported by the MEXT/JSPS KAKENHI Grant Numbers  17K05398, 18H05222, and 20H05845.
AG acknowledges support from grant numbers AST2008101 and CAREER2142300. MCHW thanks the Undergraduate Research class for their interest and participation in this research. 
Data analysis was in part carried out on the Multi-wavelength Data Analysis System operated by the Astronomy Data Center (ADC), National Astronomical Observatory of Japan.

\vspace{5mm}
\facilities{ALMA, VLA, e-MERLIN}

\software{Miriad \citep{Sault1995}}

\bibliographystyle{aasjournal}
\bibliography{SrcI.bib}

\begin{thebibliography}{}
\expandafter\ifx\csname natexlab\endcsname\relax\def\natexlab#1{#1}\fi
\providecommand{\url}[1]{\href{#1}{#1}}

\bibitem[{{Anglada} {et~al.}(2018){Anglada}, {Rodr{\'\i}guez}, \&
  {Carrasco-Gonz{\'a}lez}}]{Anglada2018}
{Anglada}, G., {Rodr{\'\i}guez}, L.~F., \& {Carrasco-Gonz{\'a}lez}, C. 2018,
  \aapr, 26, 3

\bibitem[{{B{\'a}ez-Rubio} {et~al.}(2018){B{\'a}ez-Rubio}, {Jim{\'e}nez-Serra},
  {Mart{\'\i}n-Pintado}, {Zhang}, \& {Curiel}}]{Baez-Rubio2018}
{B{\'a}ez-Rubio}, A., {Jim{\'e}nez-Serra}, I., {Mart{\'\i}n-Pintado}, J.,
  {Zhang}, Q., \& {Curiel}, S. 2018, \apj, 853, 4

\bibitem[{{Bally} {et~al.}(2020){Bally}, {Ginsburg}, {Forbrich}, \&
  {Vargas-Gonz{\'a}lez}}]{Bally2020}
{Bally}, J., {Ginsburg}, A., {Forbrich}, J., \& {Vargas-Gonz{\'a}lez}, J. 2020,
  \apj, 889, 178

\bibitem[{{Blandford} \& {Payne}(1982)}]{Blandford1982}
{Blandford}, R.~D., \& {Payne}, D.~G. 1982, \mnras, 199, 883

\bibitem[{{Commer{\c{c}}on} {et~al.}(2022){Commer{\c{c}}on}, {Gonz{\'a}lez},
  {Mignon-Risse}, {Hennebelle}, \& {Vaytet}}]{Commercon2022}
{Commer{\c{c}}on}, B., {Gonz{\'a}lez}, M., {Mignon-Risse}, R., {Hennebelle},
  P., \& {Vaytet}, N. 2022, \aap, 658, A52

\bibitem[{{Cunningham} {et~al.}(2009){Cunningham}, {Moeckel}, \&
  {Bally}}]{Cunningham2009}
{Cunningham}, N.~J., {Moeckel}, N., \& {Bally}, J. 2009, \apj, 692, 943

\bibitem[{{Curiel} {et~al.}(2006){Curiel}, {Ho}, {Patel}, {Torrelles},
  {Rodr{\'\i}guez}, {Trinidad}, {Cant{\'o}}, {Hern{\'a}ndez}, {G{\'o}mez},
  {Garay}, \& {Anglada}}]{Curiel2006}
{Curiel}, S., {Ho}, P.~T.~P., {Patel}, N.~A., {et~al.} 2006, \apj, 638, 878

\bibitem[{Forbrich {et~al.}(2021)Forbrich, Dzib, Reid, \&
  Menten}]{Forbrich_2021}
Forbrich, J., Dzib, S.~A., Reid, M.~J., \& Menten, K.~M. 2021, The
  Astrophysical Journal, 906, 23.
\newblock \url{https://doi.org/10.3847/1538-4357/abc68e}

\bibitem[{{Forbrich} {et~al.}(2016){Forbrich}, {Rivilla}, {Menten}, {Reid},
  {Chandler}, {Rau}, {Bhatnagar}, {Wolk}, \& {Meingast}}]{Forbrich2016}
{Forbrich}, J., {Rivilla}, V.~M., {Menten}, K.~M., {et~al.} 2016, \apj, 822, 93

\bibitem[{{Ginsburg} {et~al.}(2018){Ginsburg}, {Bally}, {Goddi}, {Plambeck}, \&
  {Wright}}]{Ginsburg2018}
{Ginsburg}, A., {Bally}, J., {Goddi}, C., {Plambeck}, R., \& {Wright}, M. 2018,
  \apj, 860, 119

\bibitem[{Ginsburg {et~al.}(2019)Ginsburg, McGuire, Plambeck, Bally, Goddi, \&
  Wright}]{Ginsburg2019}
Ginsburg, A., McGuire, B., Plambeck, R., {et~al.} 2019, The Astrophysical
  Journal, 872, 54.
\newblock \url{https://doi.org/10.3847%2F1538-4357%2Faafb71}

\bibitem[{{Goddi} {et~al.}(2009){Goddi}, {Greenhill}, {Chandler}, {Humphreys},
  {Matthews}, \& {Gray}}]{Goddi2009}
{Goddi}, C., {Greenhill}, L.~J., {Chandler}, C.~J., {et~al.} 2009, \apj, 698,
  1165

\bibitem[{{Goddi} {et~al.}(2011{\natexlab{a}}){Goddi}, {Greenhill},
  {Humphreys}, {Chandler}, \& {Matthews}}]{Goddi2011b}
{Goddi}, C., {Greenhill}, L.~J., {Humphreys}, E.~M.~L., {Chandler}, C.~J., \&
  {Matthews}, L.~D. 2011{\natexlab{a}}, \apjl, 739, L13

\bibitem[{{Goddi} {et~al.}(2011{\natexlab{b}}){Goddi}, {Humphreys},
  {Greenhill}, {Chandler}, \& {Matthews}}]{Goddi2011}
{Goddi}, C., {Humphreys}, E.~M.~L., {Greenhill}, L.~J., {Chandler}, C.~J., \&
  {Matthews}, L.~D. 2011{\natexlab{b}}, \apj, 728, 15

\bibitem[{{G{\'o}mez} {et~al.}(2008){G{\'o}mez}, {Rodr{\'{\i}}guez}, {Loinard},
  {Lizano}, {Allen}, {Poveda}, \& {Menten}}]{Gomez2008}
{G{\'o}mez}, L., {Rodr{\'{\i}}guez}, L.~F., {Loinard}, L., {et~al.} 2008, \apj,
  685, 333

\bibitem[{{Gordon} \& {Sorochenko}(2002)}]{Gordon2002}
{Gordon}, M.~A., \& {Sorochenko}, R.~L. 2002, {Radio Recombination Lines. Their
  Physics and Astronomical Applications}, Vol. 282,
  doi:10.1007/978-0-387-09604-9

\bibitem[{{Greenhill} {et~al.}(2013){Greenhill}, {Goddi}, {Chandler},
  {Matthews}, \& {Humphreys}}]{Greenhill2013}
{Greenhill}, L.~J., {Goddi}, C., {Chandler}, C.~J., {Matthews}, L.~D., \&
  {Humphreys}, E.~M.~L. 2013, \apjl, 770, L32

\bibitem[{{Hirota} {et~al.}(2016){Hirota}, {Kim}, \& {Honma}}]{Hirota2016a}
{Hirota}, T., {Kim}, M.~K., \& {Honma}, M. 2016, \apj, 817, 168

\bibitem[{{Hirota} {et~al.}(2014){Hirota}, {Kim}, {Kurono}, \&
  {Honma}}]{Hirota2014}
{Hirota}, T., {Kim}, M.~K., {Kurono}, Y., \& {Honma}, M. 2014, \apjl, 782, L28

\bibitem[{{Hirota} {et~al.}(2015){Hirota}, {Kim}, {Kurono}, \&
  {Honma}}]{Hirota2015}
---. 2015, \apj, 801, 82

\bibitem[{{Hirota} {et~al.}(2017){Hirota}, {Machida}, {Matsushita}, {Motogi},
  {Matsumoto}, {Kim}, {Burns}, \& {Honma}}]{Hirota2017}
{Hirota}, T., {Machida}, M.~N., {Matsushita}, Y., {et~al.} 2017, Nature
  Astronomy, 1, 0146

\bibitem[{{Hirota} {et~al.}(2020){Hirota}, {Plambeck}, {Wright}, {Machida},
  {Matsushita}, {Motogi}, {Kim}, {Burns}, \& {Honma}}]{Hirota2020}
{Hirota}, T., {Plambeck}, R.~L., {Wright}, M. C.~H., {et~al.} 2020, \apj, 896,
  157

\bibitem[{{Hoang}(2021)}]{Hoang2021}
{Hoang}, T. 2021, \apj, 921, 21

\bibitem[{{Hoang} {et~al.}(2019){Hoang}, {Tram}, {Lee}, \& {Ahn}}]{Hoang2019}
{Hoang}, T., {Tram}, L.~N., {Lee}, H., \& {Ahn}, S.-H. 2019, Nature Astronomy,
  3, 766

\bibitem[{{Hosokawa} \& {Omukai}(2009)}]{Hosokawa2009}
{Hosokawa}, T., \& {Omukai}, K. 2009, \apj, 691, 823

\bibitem[{{Issaoun} {et~al.}(2017){Issaoun}, {Goddi}, {Matthews}, {Greenhill},
  {Gray}, {Humphreys}, {Chand ler}, {Krumholz}, \& {Falcke}}]{Issaoun2017}
{Issaoun}, S., {Goddi}, C., {Matthews}, L.~D., {et~al.} 2017, \aap, 606, A126

\bibitem[{Jimenez-Serra {et~al.}(2011)Jimenez-Serra, Mart{\'\i}n-Pintado,
  Baez-Rubio, Patel, \& Thum}]{Jimenez-Serra2011}
Jimenez-Serra, I., Mart{\'\i}n-Pintado, J., Baez-Rubio, A., Patel, N., \& Thum,
  C. 2011, The Astrophysical Journal Letters, 732, L27

\bibitem[{{Jones} {et~al.}(1996){Jones}, {Tielens}, \&
  {Hollenbach}}]{Jones1996}
{Jones}, A.~P., {Tielens}, A.~G.~G.~M., \& {Hollenbach}, D.~J. 1996, \apj, 469,
  740

\bibitem[{{Kim} {et~al.}(2008){Kim}, {Hirota}, {Honma}, {Kobayashi},
  {Bushimata}, {Choi}, {Imai}, {Iwadate}, {Jike}, {Kameno}, {Kameya},
  {Kamohara}, {Kan-Ya}, {Kawaguchi}, {Kuji}, {Kurayama}, {Manabe}, {Matsui},
  {Matsumoto}, {Miyaji}, {Nagayama}, {Nakagawa}, {Oh}, {Omodaka}, {Oyama},
  {Sakai}, {Sasao}, {Sato}, {Sato}, {Shibata}, {Tamura}, \&
  {Yamashita}}]{Kim2008}
{Kim}, M.~K., {Hirota}, T., {Honma}, M., {et~al.} 2008, \pasj, 60, 991

\bibitem[{{Kounkel} {et~al.}(2018){Kounkel}, {Covey}, {Su{\'a}rez},
  {Hernandez}, {Stassun}, {Jaehnig}, {Feigelson}, {Pe{\~n}a Ram{\'\i}rez},
  {Roman-Lopes}, {Da Rio}, {Stringfellow}, {Kim}, {Borissova},
  {Fern{\'a}ndez-Trincado}, {Burgasser}, {Garc{\'\i}a-Hern{\'a}ndez}, {Zamora},
  {Pan}, \& {Nitschelm}}]{Kounkel2018}
{Kounkel}, M., {Covey}, K., {Su{\'a}rez}, Genaro a
  nd~{Rom{\'a}n-Z{\'u}{\~n}iga}, C., {et~al.} 2018, \aj, 156, 84

\bibitem[{Mariñoso~Guiu {et~al.}(2021)Mariñoso~Guiu, Ferrero,
  Macià~Escatllar, Rimola, \& Bromley}]{Guiu2021}
Mariñoso~Guiu, J., Ferrero, S., Macià~Escatllar, A., Rimola, A., \& Bromley,
  S.~T. 2021, Frontiers in Astronomy and Space Sciences, 8,
  doi:10.3389/fspas.2021.676548.
\newblock \url{https://www.frontiersin.org/articles/10.3389/fspas.2021.676548}

\bibitem[{{Matsushita} {et~al.}(2018){Matsushita}, {Sakurai}, {Hosokawa}, \&
  {Machida}}]{Matsushita2018}
{Matsushita}, Y., {Sakurai}, Y., {Hosokawa}, T., \& {Machida}, M.~N. 2018,
  \mnras, 475, 391

\bibitem[{{Matthews} {et~al.}(2010){Matthews}, {Greenhill}, {Goddi}, {Chand
  ler}, {Humphreys}, \& {Kunz}}]{Matthews2010}
{Matthews}, L.~D., {Greenhill}, L.~J., {Goddi}, C., {et~al.} 2010, \apj, 708,
  80

\bibitem[{{Menten} {et~al.}(2007){Menten}, {Reid}, {Forbrich}, \&
  {Brunthaler}}]{Menten2007}
{Menten}, K.~M., {Reid}, M.~J., {Forbrich}, J., \& {Brunthaler}, A. 2007, \aap,
  474, 515

\bibitem[{{Niederhofer} {et~al.}(2012){Niederhofer}, {Humphreys}, \&
  {Goddi}}]{Niederhofer2012}
{Niederhofer}, F., {Humphreys}, E.~M.~L., \& {Goddi}, C. 2012, \aap, 548, A69

\bibitem[{{Plambeck} \& {Wright}(2016)}]{Plambeck2016}
{Plambeck}, R.~L., \& {Wright}, M.~C.~H. 2016, \apj, 833, 219

\bibitem[{{Plambeck} {et~al.}(2009){Plambeck}, {Wright}, {Friedel}, {Widicus
  Weaver}, {Bolatto}, {Pound}, {Woody}, {Lamb}, \& {Scott}}]{Plambeck2009}
{Plambeck}, R.~L., {Wright}, M.~C.~H., {Friedel}, D.~N., {et~al.} 2009, \apjl,
  704, L25

\bibitem[{{Purser} {et~al.}(2016){Purser}, {Lumsden}, {Hoare}, {Urquhart},
  {Cunningham}, {Purcell}, {Brooks}, {Garay}, {G{\'u}zman}, \&
  {Voronkov}}]{Purser2016}
{Purser}, S.~J.~D., {Lumsden}, S.~L., {Hoare}, M.~G., {et~al.} 2016, \mnras,
  460, 1039

\bibitem[{{Rafikov}(2006)}]{Rafikov2006}
{Rafikov}, R.~R. 2006, \apj, 646, 288

\bibitem[{{Reid} {et~al.}(2007){Reid}, {Menten}, {Greenhill}, \&
  {Chandler}}]{Reid2007}
{Reid}, M.~J., {Menten}, K.~M., {Greenhill}, L.~J., \& {Chandler}, C.~J. 2007,
  \apj, 664, 950

\bibitem[{{Reynolds}(1986)}]{Reynolds1986}
{Reynolds}, S.~P. 1986, \apj, 304, 713

\bibitem[{{Rodr{\'{\i}}guez} {et~al.}(2017){Rodr{\'{\i}}guez}, {Dzib},
  {Loinard}, {Zapata}, {G{\'o}mez}, {Menten}, \& {Lizano}}]{Rodgriguez2017}
{Rodr{\'{\i}}guez}, L.~F., {Dzib}, S.~A., {Loinard}, L., {et~al.} 2017, \apj,
  834, 140

\bibitem[{{Rodr{\'{\i}}guez} {et~al.}(2005){Rodr{\'{\i}}guez}, {Poveda},
  {Lizano}, \& {Allen}}]{Rodriguez2005}
{Rodr{\'{\i}}guez}, L.~F., {Poveda}, A., {Lizano}, S., \& {Allen}, C. 2005,
  \apjl, 627, L65

\bibitem[{Rodríguez {et~al.}(2017)Rodríguez, Dzib, Loinard, Zapata, Gómez,
  Menten, \& Lizano}]{Rodriguez2017}
Rodríguez, L.~F., Dzib, S.~A., Loinard, L., {et~al.} 2017, The Astrophysical
  Journal, 834, 140.
\newblock \url{https://dx.doi.org/10.3847/1538-4357/834/2/140}

\bibitem[{Rodríguez {et~al.}(2020)Rodríguez, Dzib, Zapata, Lizano, Loinard,
  Menten, \& Gómez}]{Rodriguez2020}
Rodríguez, L.~F., Dzib, S.~A., Zapata, L., {et~al.} 2020, The Astrophysical
  Journal, 892, 82.
\newblock \url{https://dx.doi.org/10.3847/1538-4357/ab7816}

\bibitem[{{Sanna, A.} {et~al.}(2018){Sanna, A.}, {Moscadelli, L.}, {Goddi, C.},
  {Krishnan, V.}, \& {Massi, F.}}]{Sanna2018}
{Sanna, A.}, {Moscadelli, L.}, {Goddi, C.}, {Krishnan, V.}, \& {Massi, F.}
  2018, A\&A, 619, A107.
\newblock \url{https://doi.org/10.1051/0004-6361/201833573}

\bibitem[{{Sault} {et~al.}(1995){Sault}, {Teuben}, \& {Wright}}]{Sault1995}
{Sault}, R.~J., {Teuben}, P.~J., \& {Wright}, M.~C.~H. 1995, Astronomical
  Society of the Pacific Conference Series, Vol.~77, {A Retrospective View of
  MIRIAD}, ed. R.~A. {Shaw}, H.~E. {Payne}, \& J.~J.~E. {Hayes}, 433

\bibitem[{{Shull} \& {McKee}(1979)}]{Shull1979}
{Shull}, J.~M., \& {McKee}, C.~F. 1979, \apj, 227, 131

\bibitem[{Tachibana {et~al.}(2019)Tachibana, Kamizuka, Hirota, mi~Sakai, Oya,
  Takigawa, \& Yamamoto}]{Tachibana2019}
Tachibana, S., Kamizuka, T., Hirota, T., {et~al.} 2019, The Astrophysical
  Journal, 875, L29.
\newblock \url{https://doi.org/10.3847%2F2041-8213%2Fab1653}

\bibitem[{{Tanaka} {et~al.}(2020){Tanaka}, {Zhang}, {Hirota}, {Sakai},
  {Motogi}, {Tomida}, {Tan}, {Rosero}, {Higuchi}, {Ohashi}, {Liu}, \&
  {Sugiyama}}]{Tanaka2020}
{Tanaka}, K. E.~I., {Zhang}, Y., {Hirota}, T., {et~al.} 2020, \apjl, 900, L2

\bibitem[{{Testi} {et~al.}(2010){Testi}, {Tan}, \& {Palla}}]{Testi2010}
{Testi}, L., {Tan}, J.~C., \& {Palla}, F. 2010, \aap, 522, A44

\bibitem[{{Tram} \& {Hoang}(2019)}]{Tram2019}
{Tram}, L.~N., \& {Hoang}, T. 2019, \apj, 886, 44

\bibitem[{{Vaidya} \& {Goddi}(2013)}]{Vaidya2013}
{Vaidya}, B., \& {Goddi}, C. 2013, \mnras, 429, L50

\bibitem[{{Wilson} {et~al.}(2009){Wilson}, {Rohlfs}, \&
  {H{\"u}ttemeister}}]{Wilson2009}
{Wilson}, T.~L., {Rohlfs}, K., \& {H{\"u}ttemeister}, S. 2009, {Tools of Radio
  Astronomy}, doi:10.1007/978-3-540-85122-6

\bibitem[{{Wright} {et~al.}(2020){Wright}, {Plambeck}, {Hirota}, {Ginsburg},
  {McGuire}, {Bally}, \& {Goddi}}]{Wright2020}
{Wright}, M., {Plambeck}, R., {Hirota}, T., {et~al.} 2020, \apj, 889, 155

\bibitem[{{Wright} {et~al.}(2022){Wright}, {Bally}, {Hirota}, {Miller},
  {Harding}, {Colleluori}, {Ginsburg}, {Goddi}, \& {McGuire}}]{Wright2022}
{Wright}, M., {Bally}, J., {Hirota}, T., {et~al.} 2022, \apj, 924, 107

\end{thebibliography}

\end{document}